\documentclass[preprint,12pt]{elsarticle}
\usepackage{amssymb}
\usepackage{epsfig}
\usepackage{amsthm}
\usepackage{amsmath}
\journal{Physica A}
\begin{document}
\begin{frontmatter}

\title{Understanding correlation between two bound state electrons using a simple model}
\author{Shivani Verma and Aniruddha Chakraborty}
\address{Indian Institute of Technology Mandi, Kamand, Himachal Pradesh - 175075, India.}
\cortext[cor1]{Corresponding author}
\fntext[fn1]{Phone:-+91-XXX}

\begin{abstract} 
Understanding electron correlation requires solving inseparable Schrodinger equation. In general, inseparable Schrödinger equations cannot be solved analytically. So their solutions are obtained numerically. In this paper we investigate electron correlation problem using the Dirac delta function repulsion between two electrons, where each electron is bound in their respective Dirac delta potential well. 
\end{abstract}

\begin{keyword} Electron correlation; Bound state; Dirac delta.
\end{keyword}\end{frontmatter}
\noindent In recent years it has become increasingly evident that understanding electron correlation better is very important \cite{Raghavachari}. There are several different models to explain electron correlation in two electron systems \cite{hylleraas, Analytic approach}. All of these methods are approximate method. Most simplest model for understanding electron correlation is the case of two electron system \cite{Two deltons}. The most simplest possible Hamiltonian for a system of two interacting electrons may be written as follows
\begin{equation}
\hat{H}= \frac{-\hbar^2}{2m}\left(\frac{\partial^2}{\partial x^2}+\frac{\partial^2}{\partial y^2}\right)-\alpha \delta(x)- \alpha \delta(y)+ \lambda\delta(x-y),
\end{equation}
\noindent where coordinate of one electron is denoted by $x$ and the coordinate of other electron is represented by $y$ and $m$ denote the mass of the electron. Both the electrons are trapped in a Dirac delta potential well and electron-electron repulsion term is approximated by adding one Dirac delta potential barrier between the electrons and is denoted by $\delta(x-y)$. The corresponding time independent Schrodinger equation is given by
\begin{eqnarray} \label{eq:3.5}
\frac{-\hbar^2}{2m}\left(\frac{\partial^2}{\partial x^2}+\frac{\partial^2}{\partial y^2} \right)\psi(x,y)-\alpha \delta(x)\psi(x,y)-\alpha \delta(y)\psi(x,y)
+\lambda \delta(x-y)\psi(x,y)=E\psi(x,y)
\end{eqnarray}
\noindent In the region where, $x \neq 0$, $y \neq 0$ and $x \neq y$, then, the above equation becomes,
\begin{equation}\label{eq:3.9}
   \bigg(\frac{\partial^2}{\partial x^2}+\frac{\partial^2}{\partial y^2}\bigg)\psi(x,y)=k^2  \psi(x,y)    
\end{equation}
\noindent where $k^2$ is a positive constant and is given by
\begin{equation} \label{eq:3.8}
k^2 = \frac{-2mE}{\hbar^2}.  
\end{equation}
\noindent In the above we assumed $E$ to be negative, as we are dealing with bound states. Now Eq. (\ref{eq:3.9}) can be solved by using method of separation of variables as given below
\begin{equation}
\psi(x,y)=\phi(x)\phi(y)\label{eq:3.10}
\end{equation}
\par Substituting eq. (\ref{eq:3.9}) in eq. (\ref{eq:3.10}) and dividing the resulting equation by $\phi(x)\phi(y)$, we get
\begin{equation}
\frac{1}{\phi(x)}\frac{\partial^2}{\partial x^2} \phi(x)+\frac{1}{\phi(y)}\frac{\partial^2}{\partial
y^2}\phi(y)=k^2 
\end{equation}
\noindent In the above equation first part of the L.H.S. is a function of \textit{x} only and second  part of the L.H.S. is the function of \textit{y} only but their summation is a constant {\it i.e.,} $k^2$. It is possible only when the each part of the L.H.S. is a separate constant. Since, first part and second part of the L.H.S. are identical. So, we can split the equation in two parts and both the parts will be equal to $\frac{k^2}{2}$.
\begin{eqnarray} \label{eq:3.14}
   \frac{1}{\phi(x)}\frac{\partial^2}{\partial x^2}\phi(x)=\frac{k^2}{2} \\ \nonumber
   \frac{1}{\phi(y)}\frac{\partial^2}{\partial y^2}\phi(y)=\frac{k^2}{2}
\end{eqnarray}
\noindent On further simplification, we get
\begin{eqnarray}
 \frac{\partial^2}{\partial x^2}\phi(x)=\frac{k^2}{2}\phi(x)\\ \nonumber
\frac{\partial^2}{\partial y^2}\phi(y)=\frac{k^2}{2}\phi(y).
\end{eqnarray}
\noindent
\par On solving Eq. (8) we get the following possible solutions
\begin{equation} \label{eq:3.16}
  \phi(x)=Ae^\frac{-kx}{\sqrt{2}}\qquad \& \qquad \phi(x)=Ae^\frac{kx}{\sqrt{2}} 
  \end{equation}
  \begin{equation} \label{eq:3.17}
  \phi(y)=Ae^\frac{-ky}{\sqrt{2}} \qquad \& \qquad \phi(y)=Ae^\frac{ky}{\sqrt{2}}   
  \end{equation}
\noindent where $A$ is an unknown constant, to be determined later. As we are dealing with bound states, so the acceptable wave-function should vanish at infinity and it has to be continuous at all points. So the analytical formula of acceptable wave-function is given by
\begin{eqnarray}
   \phi(x)=Ae^\frac{-k|x|}{\sqrt{2}} \\ \nonumber
   \phi(y)=Ae^\frac{-k|y|}{\sqrt{2}} 
\end{eqnarray}
\noindent Therefore, the total wave function is given by
\begin{equation}\label{eq:3.21}
\psi(x,y)=A e^\frac{-k(|x|+|y|)}{\sqrt{2}}   
\end{equation}
\noindent Now we ask the question, is the wave-function given above is the only acceptable wave-function. To understand that we should look at all possible combinations of wave-functions given in Eq. (9) and Eq. (10). We get four possible options, which are listed below
\begin{eqnarray} \label{eq:3.22}
         \psi(x,y)=A\;e^\frac{-kx}{\sqrt{2}}\;e^\frac{-ky}{\sqrt{2}}\\    
         \psi(x,y)=A\; e^\frac{kx}{\sqrt{2}}\;e^\frac{ky}{\sqrt{2}}  \\ 
         \psi(x,y)=A\; e^\frac{-kx}{\sqrt{2}}\;e^\frac{ky}{\sqrt{2}}  \\ 
         \psi(x,y)=A\; e^\frac{kx}{\sqrt{2}}\;e^\frac{-ky}{\sqrt{2}}   
\end{eqnarray}
\noindent Now we will analyze the above four options to see the possibility of getting any more acceptable wave-function other than the one in Eq. (12). Actually the solutions given in Eq. (13) and Eq. (14) is used in deriving Eq.(12). So now focus should be on Eq. (15) and Eq. (16).
\begin{figure}[ht]
\centering
\includegraphics[width=0.5\textwidth]{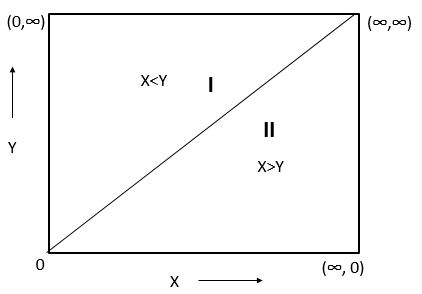}
\caption{Coordinate system describing the location of electron 1 and 2.}
\label{fig:mesh3}
\end{figure}
\noindent For region I, where $\textit{x}<\textit{y}$, so that $|x-y| = - (x-y)$ the physically acceptable wave-function is the one that is given in  Eq. (16). Again for region II, where $\textit{x}>\textit{y}$, so that $|x-y| =  (x-y)$ the physically acceptable wave-function is the one that is given by Eq. (15). So the physically acceptable solution for all regions can be given by combining Eq.(15) and Eq.(16), as given below
\begin{equation} \label{eq:3.26}
\psi(x,y)= B e^\frac{-k|x-y|}{\sqrt{2}}   
\end{equation}
\noindent So the better wave-function for two electron system can be represented as a linear combination of wave-function which we get from Eq. (\ref{eq:3.21}) and Eq. (\ref{eq:3.26}), which is given by
\begin{equation}
\psi(x,y)=A e^\frac{-k(|x|+|y|)}{\sqrt{2}}+ B e^\frac{-k|x-y|}{\sqrt{2}}   
\end{equation}
\noindent Since, the Hamiltonian involved three Dirac delta functions. Each Dirac delta function will lead to two boundary conditions for wave-functions. In the following we will derive all these boundary conditions. First we Integrate Eq. (\ref{eq:3.5}) w.r.t. $x$ from $0-\epsilon$ to $0+\epsilon$, under the limit  $\epsilon\rightarrow 0$. We get
\begin{equation} \label{eq:3.28}
\frac{-\hbar^2}{2m}\left[\frac{\partial\psi(x,y)}{\partial x}\right]^{0+\epsilon}_{0-\epsilon} -\alpha \psi(0,y)+\lambda \psi(y,y)=0
\end{equation}
\noindent Now we put $y=0$, so that we get boundary condition at ($x=0, \; y=0$).
\begin{equation}
\frac{-\hbar^2}{2m}\left[\frac{\partial\psi(x,y)}{\partial x}\right]^{0+\epsilon}_{0-\epsilon} -\alpha \psi(0,0)+\lambda \psi(0,0)=0
\end{equation}
\noindent This is first boundary condition. Now we integrate Eq. (\ref{eq:3.5}) w.r.t. $y$ from $0-\epsilon$ to $0+\epsilon$  under the limit  $\epsilon \rightarrow 0$ to get
\begin{equation} \label{eq:3.30}
\frac{-\hbar^2}{2m}\left[\frac{\partial\psi(x,y)}{\partial y}\right]^{0+\epsilon}_{0-\epsilon} -\alpha \psi(x,0)+\lambda \psi(x,x)=0
\end{equation}
\noindent Now we put $y=0$, so that we get boundary condition at ($x=0, \; y=0$).
\begin{equation}
\frac{-\hbar^2}{2m}\left[\frac{\partial\psi(x,y)}{\partial y}\right]^{0+\epsilon}_{0-\epsilon} -\alpha \psi(0,0)+\lambda \psi(0,0)=0
\end{equation}
\noindent  Now we integrate Eq. (\ref{eq:3.5}) w.r.t. $x$ from $y-\epsilon$ to $y+\epsilon$ (with the assumption  $\textit{y}\neq 0$) under the limit $\epsilon \rightarrow 0$ to get
\begin{equation}
\frac{-\hbar^2}{2m}\left[\frac{\partial\psi(x,y)}{\partial x}\right]^{y+\epsilon}_{y-\epsilon}+\lambda \psi(y,y)=0.
\end{equation}
\noindent Now we integrate Eq. (\ref{eq:3.5}) w.r.t. $y$ from $x-\epsilon$ to $x+\epsilon$ (with the assumption  $\textit{x}\neq 0$) under the limit $\epsilon \rightarrow 0$ to get
\begin{equation}
\frac{-\hbar^2}{2m}\left[\frac{\partial\psi(x,y)}{\partial y}\right]^{x+\epsilon}_{x-\epsilon} +\lambda \psi(x,x)=0
\end{equation} 
\noindent If we assume $\textit{y}\neq 0$, then Eq. (\ref{eq:3.28}) reduces to,
\begin{equation} \label{eq:3.34}
\frac{-\hbar^2}{2m}\left[\frac{\partial\psi(x,y)}{\partial x}\right]^{0+\epsilon}_{0-\epsilon} -\alpha \psi(0,y)=0
\end{equation}
\noindent If we assume $\textit{x}\neq 0 $ then  Eq. (\ref{eq:3.30}) reduces to,
\begin{equation} \label{eq:3.35}
\frac{-\hbar^2}{2m}\left[\frac{\partial\psi(x,y)}{\partial y}\right]^{0+\epsilon}_{0-\epsilon} -\alpha \psi(x,0)=0
\end{equation}
\noindent Therefore, we have all six boundary conditions. In  our model $\lambda$ is the strength by which the two electrons interacting with each other, where $\lambda=0$ is for completely uncorrelated system and $\lambda=1$ for completely correlated system. Assuming this model is true for completely correlated system (so that $\lambda = 1$) and taking $\alpha=1$ for simplicity. Then, the first two boundary conditions will be of no use because they will not have any electron-correlation effect. Now, the six boundary conditions will be reduced to only four boundary conditions which are as follows:
\begin{equation} 
\label{eq:3.36}
\frac{-\hbar^2}{2m}\left[\frac{\partial\psi(x,y)}{\partial x}\right]^{y+\epsilon}_{y-\epsilon}+\psi(y,y)=0
,\qquad  \textit{y}\neq0
\end{equation}

\begin{equation} \label{eq:3.37}
\frac{-\hbar^2}{2m}\left[\frac{\partial\psi(x,y)}{\partial y}\right]^{x+\epsilon}_{x-\epsilon} +\psi(x,x)=0,\qquad
\textit{x}\neq0
\end{equation}

\begin{equation} \label{eq:3.38}
\frac{-\hbar^2}{2m}\left[\frac{\partial\psi(x,y)}{\partial x}\right]^{0+\epsilon}_{0-\epsilon} - \psi(0,y)=0,\qquad\textit{y}\neq0
\end{equation}
\begin{equation} \label{eq:3.39}
\frac{-\hbar^2}{2m}\left[\frac{\partial\psi(x,y)}{\partial y}\right]^{0+\epsilon}_{0-\epsilon} -\psi(x,0)=0
,\qquad\textit{x}\neq0
\end{equation}
\noindent The above four boundary conditions can be use to calculate the coefficients in the wave-function as well as the total energy of the system. On solving the Eq. (\ref{eq:3.36}) and Eq. (\ref{eq:3.37}) we get
\begin{equation} 
\label{eq:3.40}
\frac{\hbar^2}{m} \frac{k}{\sqrt{2}} B + B+ A e^{-k\sqrt{2}x}=0
\end{equation}
\begin{equation} \label{eq:3.41}
    \frac{\hbar^2}{m} \frac{k}{\sqrt{2}} B + B+ A e^{-k\sqrt{2}y}=0
\end{equation}
\noindent On solving eq (\ref{eq:3.38}) and (\ref{eq:3.39}) we get
\begin{equation} \label{eq:3.42}
    A \bigg(-\frac{k}{\sqrt{2}}+\frac{m}{\hbar^2}\bigg)
+ B \frac{m}{\hbar^2}=0
\end{equation}
\noindent For electron 1  we can take eq (\ref{eq:3.40}) and (\ref{eq:3.42}) and solve them. These equations may have trivial solution, if the determinant formed by the matrix (say P) of above equation is zero. So, we will not consider those trivial solutions here, where, $|P| \neq 0 $. But, if $|P|=0$ Then, we will have non-trivial solutions.
\begin{center}
\par If $x=0$  \qquad $|P|=0$
\par If $x\neq0$  \qquad $|P|\neq0$\\
\end{center}
\noindent Using this property of linear homogeneous equation \cite{mort}, we can determine the relation between $k$ and $x$, which is
\begin{equation} \label{eq:3.43}
\frac{m}{\hbar^2} - \frac{\hbar^2}{m}\frac{k^2}{2}- \frac{m}{\hbar^2} e^{-k\sqrt{2} x } =0 
\end{equation}
In atomic units, mass of electron is unity and $\hbar$ is also $1$. Therefore Eq. (34) becomes
\begin{equation} \label{eq:3.44}
1-\frac{k^2}{2}-e^{-k\sqrt{2}x}=0
\end{equation}
This Eq. (\ref{eq:3.44}) and Eq. (\ref{eq:4.6}) gives data for $k$, position and energy of electron.
\begin{center}
 \begin{tabular}{||c| c| c ||} 
 \hline
 x & k & E(in atomic units) \\ [0.5ex] 
 \hline\hline
 1 & 1.29622 & -0.840093  \\ 
 \hline
 2 & 1.40069 & -0.980966  \\
 \hline
 3 & 1.41245 & -0.995507  \\
 \hline
 4 & 1.41398 & -0.999669  \\
 \hline
 5 & 1.41418 & -0.999952  \\ 
 \hline
  6 & 1.41421 & -0.999994  \\
 \hline
  7 & 1.41421 & -0.999994  \\
 \hline
  8 & 1.41421 & -0.999994  \\
 \hline 
 9 & 1.41421 & -0.999994  \\
 \hline
 10  & 1.41421 & -0.999994  \\ [1ex]
 \hline  
\end{tabular}
\end{center}
\noindent According to this equation, $k$ can be plotted as function of $x$,
\begin{figure}[h]
\centering
\includegraphics[width=0.6\textwidth]{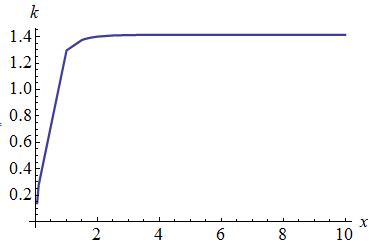}
\caption{Variation of k with the distance of electron 1 from the nucleus}
\label{fig:6}
\end{figure}
\noindent Firstly, as the distance between the electron and nucleus increases the value of $k$ also increases but after a while on increasing the distance $k$ remains constant.
\par
\begin{figure}[ht]
\centering
\includegraphics[width=0.6\textwidth]{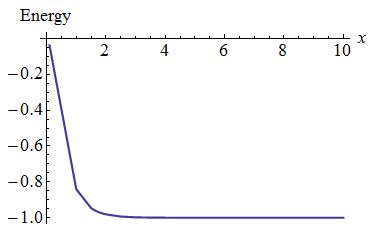}
\caption{Variation of Energy with the distance of electron 1 from the nucleus}
\label{fig:7}
\end{figure}

 From eq (\ref{eq:3.8}) (in atomic units),
\begin{equation} \label{eq:4.6}
E=-\frac{k^2}{2}
\end{equation}
\noindent By this relation, plot {\it i.e.,} fig(\ref{fig:7}) between $E$ and $x$ is made for electron 1. Similar type of plot can be made for other electron also.
\noindent This graph shows that initially energy was decreases as distance between electron and nucleus increases and after a certain period of time energy become constant even after increasing the distance.

\section*{Summary and Conclusions}

\noindent In summary, a simple model which can show the effect of electron correlation in two electron system has been made. The effect of electron correlation on two electron system in bound states is explained graphically and numerically. Using a method very similar to the one discussed here can be applied to $N$ electron systems as well. Arbitrary bound state potentials can be modelled by using an appropriate collection of Dirac delta potentials. The corresponding time-dependent Schr\"o dinger equation can also be solved.

\section*{Acknowledgments}
\noindent One of the authors (A.C.) acknowledges IIT Mandi for CPDA Grant.

\end{document}